\documentstyle[12pt]{article}
\begin{document}
\newcommand{\abstrait}{We study the interface between soft and hard QCD
at high energy and small momentum transfer. At LHC and SSC energies, we
find that a cutoff BFKL equation leads one to expect a measurable
perturbative component in traditionally soft processes. We show that the
total cross section could become as large as 175 mb (122 mb) and the
$\rho$ parameter 0.40 (0.25) at the SSC (LHC).}
\begin{titlepage}
\begin{flushright}
McGill/92--24\\
June 1992
\end{flushright}
\vspace{.5in}
\begin{center}
\renewcommand{\thefootnote}{*}
{\large\bf{PERTURBATIVE QCD CORRECTIONS TO THE SOFT
POMERON}\footnote{Talk presented at the SSC Physics Symposium, Madison,
Wisconsin, 13-15 April 1992 and at the MRST XIV, 7-8 May 1992,
Toronto}}
\renewcommand{\thefootnote}{\dag}

\vspace{.2in}
{J.R. Cudell\footnote{CUDELL@HEP.PHYSICS.MCGILL.CA} \\
and \\
B. Margolis \\
\it Physics Department, McGill University, Montr\'eal, P.Q. H3A 2T8,
Canada\rm}
\vspace{.7 in}

{\bf Abstract}
\end{center}
\abstrait
\end{titlepage}
\newpage
\section{Introduction}
\label{intro}

\noindent As energy increases, protons look more and more like clouds of
soft partons, so that small-x and soft physics are going to give us the
typical event of future hadron-hadron colliders. Many events will contain
``minijet" structures, scattering of soft partons will have to be modeled
in background estimates, and can be used for the detection of very heavy
particles \cite{Bialas,FAD}. A detailed understanding of the total cross
section will normalize these processes.

Soft interactions are already rather well described by several models
\cite{Donnachie,TTWu,Margolis}. However, their properties cannot be
reproduced by QCD, and perturbative attempts, although infrared finite,
have totally failed so far \cite{RossH}. So, one is lead to the conclusion
that the problem is mostly non-perturbative, and that one should consider the
perturbative calculation only after cutting it off for small gluon momenta
$k_T$: the non-perturbative models then serve as a small $k_T$ term which
we then evolve using perturbative QCD.

We limit ourselves to the most general features that one can expect from
such an evolution, and do not attempt to make an explicit model. As the
QCD equations are simpler at zero momentum transfer, we consider only the
total cross section and the ratio of real to imaginary part of the forward
scattering amplitude, the $\rho$ parameter. Even then, as the exchange will
involve at least two gluons, it is possible to demand that both have large
transverse momenta, which add up to zero. The perturbative evolution then can
lead to a ``gluon bomb" which remains dormant in the data up to present
energies, but which can bring large observable corrections at future colliders.

In the next section, we give a simple model for soft physics at $t=0$, which we
call the soft pomeron. We then briefly outline the BFKL equation \cite{BFKL}
and mention its solutions, which are very far from reproducing the data. We
then show how one can make a very general model evolving soft physics to
higher values of $\log s$ and constrain it using existing data for
$\sigma_{tot}$ and $\rho$. We then show that soft physics at the SSC and the
LHC could have a substantial perturbative component.

\section{Data: the soft pomeron}

\noindent As explained above, we shall concentrate on the hadronic amplitude
 ${\cal A}\sl(s,t=0)$ describing the elastic scattering of
pp and p$\rm\bar p$ with center-of-mass energy $\sqrt{s}$ and squared
momentum transfer $t=0$. This amplitude is known experimentally, as its
imaginary part is proportional to $s$ times the total cross section, and the
ratio of its real and imaginary parts is by definition $\rho$.

The most economical fit, inspired by Regge theory,
is a sum of two simple Regge poles:
\begin{equation}
Im\ {{\cal A}\over s}=(a\pm i b) s^{\epsilon_m}+C_0 s^{\epsilon_0}
\end{equation}
with $a$, $b$, $C_0$ constants
independent of s. The phase of the amplitude is obtained by the imposition of
crossing symmetry. The first term has a universal part ($a$) representing
$f$ and $a_2$ exchange, and a part ($b$) changing sign between p
and $\rm\bar p$
scattering, which comes from $\rho$ and $\omega$ exchange. The second term
($C_0$) is responsible for the rise in $\sigma_{tot}$ and is referred to as the
``soft pomeron''. This parametrization successfully reproduces all
available data \cite{data}, from $\sqrt{s}=$ 10 Gev
to 1800 GeV. The only failure is the UA4 value for $\rho$, which is not
reproduced by most models, and for which further confirmation seems to be
needed. The curves shown in Figure 1 result from a fit to the data of
reference \cite{data}. The best fit is for the values $\epsilon_m=-0.46$ and
$\epsilon_0=0.084$. It predicts
$\sigma_{tot}=125$ mb (107 mb), $\rho=0.13$ at the SSC (LHC).

Other parametrizations are possible, {\it e.g.} \cite{TTWu,Margolis},
and as shown by the proponents of this one \cite{Donnachie}, multiple Regge
exchanges are essential to describe the data at nonzero t. However, as
we limit ourselves here to the zero momentum transfer case for which
the corrections are small, and as this simple form is particularly
well suited for our purpose, we shall adopt it in the following as a
starting point for the QCD evolution.

\section{Theory: the hard pomeron}

\noindent In order to describe total cross sections within the context
of perturbative QCD, one can try, for $s\rightarrow\infty$, to isolate
the leading contributions and to resum them. This is made possible by
the fact that perturbative QCD is
infrared finite in the leading $\log s$ approximation and in the
colour-singlet channel. This suggests that very small momenta might not
matter, and that one could use perturbation theory.

Such a program has been developed by BFKL \cite{BFKL}. In a nutshell,
one can show that, when considering gluon diagrams only,
the amplitude is a sum of terms $T_n$ of order $(\log s)^n$
and that terms of order $(\log s)^n$ are related to terms of order
$(\log s)^{n-1}$ by an integral operator that does not depend on $n$,
and that we shall write
$\hat K$:
\begin{eqnarray}
&&T_{n+1}(s,k_T^2)=\hat K T_n(s',k_T'^2)\nonumber\\
&=&{3\alpha_S\over \pi} k_T^2
\int_{s_0}^s {ds'\over s'}\int {dk_T'^2\over
k_T'^2}[{T_n(s',k_T'^2)-T_n(s',k_T^2)\over |k_T^2-k_T'^2|}
+{T_n(s',k_T^2)\over\sqrt{k_T^2+4 k_T'^2}}]
\end{eqnarray}
this leads to:
\begin{equation}
T_{\infty}=\sum_n T_n=T_0+\hat K T_{\infty}
\end{equation}
This is the BFKL equation at $t=0$. Its extension to nonzero $t$ is known, but
too complicated to handle analytically. We limit ourselves here to the zero
momentum transfer case.

In this regime, the BFKL equation (3) possesses two classes of solutions.
First of all, at
fixed $\alpha_S$, the resummed amplitude is a Regge cut instead of a simple
pole:
$T_\infty\approx\int d\nu s^{N(\nu)}$, with a leading behaviour given by
\begin{equation}
N_{max}=1+{12 \log 2\over\pi}\alpha_S
\end{equation}
Even for a small $\alpha_S$, say of order of 0.2, this leads to a big intercept
$N_{max}\approx 1.5$. As this is much too big to accomodate the data, and as a
cut rather than a pole leads to problems with quark counting, subleading
terms were added via the running of the coupling constant. It was first
claimed that such terms would discretize the cut and turn it into a series of
poles \cite{LipatovKirchner}, but further work has shown that the cut structure
remains \cite{RossH,RossD}. However, the leading singularity is slightly
reduced, and one can derive the bound \cite{CollinsK}
\begin{equation}
N_{max}>1+{3.6\over\pi}\alpha_S
\end{equation}
Again, for values of $\alpha_S$ of the order of 0.2, this leads to an
intercept of the order of 1.23.

We thus reach a contradiction: on the one hand, the data demands that the
amplitude rises more slowly than $s^{1+\epsilon_0}$, with
$\epsilon_0<0.1$; on the other hand,
perturbative resummation leads to a power $s^{1+\epsilon_p}$, with
$\epsilon_p>1.23$. The difference between
the two is a factor 3 in the total cross section at the Tevatron. The
resolution of this problem is far from clear, and one can envisage the
implementation of some non-perturbative effects within the BFKL equation
\cite{RossH}. Rather than trying to understand $\epsilon_0$,
we shall here take a
much simpler approach, {\it i.e.}
assume a low-$k_T$, low-$s$ behaviour consistent with the
data, and see what general features its perturbative evolution might exhibit.

The idea is thus to cut off equation (3) by imposing $k_T^2>Q_0^2$, with $Q_0$
big enough for perturbation theory to apply, so that one uses the perturbative
resummation only at short distances. Furthermore, one takes $T_0\sim s^{1+
\epsilon_0}$ as the non-perturbative driving term, valid for $k_T^2<Q_0^2$.
This cutoff equation has
been recently solved by Collins and Landshoff \cite{CollinsL} in the case of
deep inelastic scattering.
Most of their results and approximations can be carried over
to the hadron-hadron scattering case,
and we shall give here the basic features of the solution in this case.

First of all, the hadronic amplitude can be thought of as the convolution of
two form factors times a resummed QCD gluonic amplitude obeying a cutoff BFKL
equation.
\begin{equation}
{\cal A}(s,t)=\int_{Q_0}^{\sqrt{s}}dk_1 {V(k_1)\over k_1^4}
\int_{Q_0}^{\sqrt{s}}dk_2 {V(k_2)\over k_2^4} T(k_1,k_2;s)
\end{equation}
$k_1$ and $k_2$ are the momenta entering the gluon ladder from either hadron,
$\sqrt{s}$ is the total energy, the two form factors
$V(k_i)$, $i$=1,2, represent the coupling of the proton to the perturbative
ladder via a non-perturbative exchange, and the $1/k_i^4$ come from the
propagators of the
external legs. $T(k_1,k_2;s)$ will obey the BFKL equation both for $k_1$ and
$k_2$, and the two independent evolutions will be related by the
driving term $T_0$ representing the 2-gluon exchange
contribution and thus proportional to $\delta(k1-k2) s^{1+\epsilon_0}$.
The next terms $T_n$ will be given by equation (2) but cut off at small
$k$:
\begin{equation}
T_n(k_T,k_2;s)=\theta(\sqrt{s}>k_T>Q_0) \hat K T_{n-1}(k_T',k_2;s')
\end{equation}
Under these assumptions, and working at fixed $\alpha_s$, one can show that
the amplitude (6) conserves the structure found in \cite{CollinsL}:
\begin{equation}
{{\cal A}\over s}=C_0 s^{\epsilon_0}+\sum_{n=1}^\infty C_n(s) s^{\epsilon_n(s)}
\end{equation}
This solution reduces to the usual solution of the BFKL equation when
$s\rightarrow\infty$ and $Q_0\rightarrow 0$. The coefficients $C_n$ depend on
the model assumed for the coupling $V(k)$ between the non-perturbative and the
perturbative physics and their $s$ dependence is a threshold effect coming
from the integration in (6). Their only general property is that they are
positive.
On the other hand, the powers $\epsilon_n(s)$ are universal functions that
depend only on $\alpha_S$ and $\sqrt{s}/Q_0$.

\section{Interplay between soft and hard QCD: a model}
As the coefficients of the series (8) are model-dependent, we do not attempt to
calculate them, but rather try to assess the constraints that present
data place on them. We shall then be able to decide whether such
perturbative effects could show up in soft physics at future colliders.
As all the $C_n$ are positive, the behaviour of the series (8) will not be
very different from that of its leading term,
and so we truncate it. We also make an educated guess for the threshold
function contained in $C_1(s)$. This does not affect our results for the
values of $Q_0$ shown here. We finally impose crossing symmetry to get
the real part of the amplitude. This gives
\begin{eqnarray}
{\tilde{\cal A}\over s}&=&C_0s^{\epsilon_0} +[c_1 (1-{Q_0\over
\sqrt{s}})^2\ \theta(\sqrt{s}-Q_0)]\ s^{\epsilon_1(s)}\\
{\cal A}(s)&=&{\tilde{\cal A}(s)}+\tilde{\cal A}(s e^{-i\pi})
\end{eqnarray}
with $c_1$ a positive constant.
To calculate $\epsilon_1(s)$ we assume that $Q_0$ is the scale of
$\alpha_S$ and
take $\Lambda_{QCD}=200$ MeV. Using the results of reference \cite{CollinsL},
we calculate the curves of Figure 2(a), for various values of the cutoff $Q_0$
and thus of $\alpha_S$. One sees that the effective power is much smaller than
its purely perturbative counterpart (4), {\it e.g.} for $Q_0$=2 GeV, the usual
estimate (4) gives $\epsilon$=0.8, whereas a cutoff equation gives values half
as big at accessible energies.

As the amplitude (9) in principle violates unitarity, we also consider its
eikonalized version to see whether unitarization can make a difference.
Note that the use of such an
eikonal formalism \cite{MargolisF} is not derived from QCD. In
fact, the BFKL equation in principle sums multi-gluon ladders in the $s$
and $t$ channels, so that in the purely perturbative case the eikonal
formalism is probably too na\"\i ve. However, in this case, it can be thought
of as an expansion in the number of form factors $V(k_1) V(k_2)$. This is
definitely not included in the BFKL equation. To the amplitude (9), we further
add the meson trajectories of (1), and proceed to fit the data.

The first obvious observation is that the extra perturbative terms do not help
the fit: due to the positivity of the $C_n$, they cannot produce a bump in
$\rho$ that would explain the UA4 measurement.
So, one gets the best fit when the new QCD terms are actually
turned off. There are two ways of turning them off: either taking the infrared
cutoff $Q_0$ to infinity, or setting the coefficient $c_1$ to zero. So one can
plot the allowed regions in the $\sqrt{c_1/C_0}$, $Q_0$ plane.
We show the $2\sigma$ allowed region in figure 2(b), which comes almost
entirely from the measured value of $\sigma_{tot}$ at the Tevatron.

One can understand the general trend of this figure as follows:
the non-perturbative term at the Tevatron is about 1 mb
below the 2$\sigma$ limit, so one can ``fill" about 1 mb at the Tevatron with
a perturbative contribution. This gives an upper bound on $c_1$. If this upper
bound is realized, then the cross section at the SSC (LHC) will roughly be
$500^{\epsilon_1}$mb ($79^{\epsilon_1}$mb). This would give about 22 mb at the
SSC, for $\epsilon_1\approx 0.5$. However, one picks up an extra contribution
from the evolution of $\epsilon_1$ between the two energies, which doubles
this estimate. As this reasoning shows, the bigger $\epsilon_1$, {\it i.e.}
the smaller $Q_0$, the more dramatic the effect. As we want to choose a $Q_0$
so that the evolution is comfortably perturbative, we pick a value of 2 GeV,
and we show the resulting curves in Figure 3 and their eikonalized version in
Figure 4. The predictions for the LHC and the SSC are given in table 1.

\begin{center}
\begin{tabular} {|l|l|l|l|l|}   \hline
collider/              & pole fit     & eikonal fit & pole fit & eikonal fit\\
quantity               & NP           &NP           &P+NP      &P+NP\\ \hline
SSC $\sigma_{tot}$(mb) &125           & 120         &175       &159 \\ \hline
LHC $\sigma_{tot}$(mb) &107           & 106         &122       &121 \\ \hline
SSC $\rho$             &0.13          & 0.12        &0.40      &0.26 \\ \hline
LHC $\rho$             &0.13          & 0.12        &0.25      &0.20 \\ \hline
\end{tabular}
\end{center}

\noindent Table 1: Allowed values of the cross section and the $\rho$
parameter. The first two columns result from the purely non-perturbative (NP)
ansatz (1), and the two last columns (P+NP) from its perturbative evolution,
with $Q_0=$ 2 GeV, see equation (9).

$\ $

Clearly, unitarization is not going to make a big difference even at
SSC energies. Its main effect will be to bring down the $\rho$ parameter at
ultra high energy $\sqrt{s}\approx 10^5$ GeV. One sees from these curves
that the perturbative component
could contribute 50 mb (15 mb) at the SSC (LHC). We emphasize that this is a
conservative estimate, based on an infrared cutoff of 2 GeV. Cutting off the
evolution when $\alpha_S\approx 1$ would give a total cross section of at least
3 b at the SSC, and be consistent with all available collider and fixed target
data!

\section{Conclusion}
We have shown that the BFKL equation can be used to evolve the soft pomeron to
higher $s$, and that perturbative effects could become measurable at the
SSC/LHC. These effects are cutoff dependent, and perturbative physics seems to
couple very weakly to the proton in the diffractive region, its coupling
strength being a few percent of that of the soft pomeron. However,
even a very weak coupling turning on at an energy of a few GeV can lead
to measurable effects at sufficiently large energy. It is known that the
pomeron couples to
quarks, and quarks to gluons. The coupling to the BFKL ladder thus
cannot be zero, and specific models can be built for it \cite{cudell}.

This contribution is genuinely new and comes entirely from a QCD analysis. One
should not be mislead by previous parton models \cite{Margolis} which, while
using a partonic picture, keep it mostly non-perturbative, replacing the small
power $s^{\epsilon_0}$ of (1) by a small power $x^{-\epsilon_0}$ in the gluon
structure function $xg(x)$. In the present model, $xg(x)$ will contain the
same powers
$\epsilon_i(Q_0/x)$ as the total cross section,
but their coefficients will in general be different from those entering
the total cross section, and the relation between them will be model dependent.

The existence of such possibilities, and the fact that very large total cross
sections are expected from the same kind of arguments that lead one to
predict a
rising cross section \cite{MargolisF,ChengW}, shows that small momentum physics
contains a wealth of open possibilities worth exploring experimentally.

\section*{Acknowledgments}
This work was supported in part by NSERC (Canada) and les fonds FCAR
(Qu\'ebec).

\section*{Figure Captions}

\noindent Figure 1: Simple pole fit (1) at zero momentum
transfer, for pp and p$\rm{\bar p}$ scattering, as indicated.
(a) shows the total
cross section and (b) the ratio of the real-to-imaginary parts of the
amplitude. The data are from reference \cite{data}.

\hfill\break
\noindent Figure 2: (a) shows the effective power of s of equation (9) that
results from a cutoff BFKL equation, for various values of the infrared cutoff
$Q_0$, as indicated next to the curves. (b) gives the $2\sigma$ allowed
regions for an amplitude given by equation (9) (hatched region) or eikonalized
(cross hatched). The abscissa is the infrared cutoff $Q_0$
and the ordinate the ratio
of the coupling strength of the perturbative ladder to
that of the soft pomeron.

\hfill\break
\noindent Figure 3: same as figure 1. The upper curve shows the
maximum values of the
cross-section (a) and of $\rho$ (b) that might result from a perturbative
evolution consistent with all present data.
\hfill\break

\noindent Figure 4: same as (3), but unitarized using the eikonal formalism.

\end{document}